\documentclass[twocolumn,prb]{revtex4-2}
\usepackage{epsfig,amssymb,amsmath}
\usepackage{graphicx}
\usepackage{color}
\usepackage[english]{babel}

\begin{document}
\preprint{APS/123-QED}
\title{ Bifurcation structure and chaos in nanomagnet coupled to Josephson junction}

\author{M. Nashaat$^{1,2}$}
\author{M. Sameh$^{1}$}
\author{A. E. Botha $^{3}$}
\author{K. V. Kulikov $^{2,4}$}
\author{Yu. M. Shukrinov$^{2,4,5}$}
\affiliation{$^1$ \mbox{Department of Physics, Faculty of Science, Cairo University, 12613, Giza, Egypt}\\
$^2$ \mbox{BLTP, JINR, Dubna, Moscow Region, 141980, Russia}\\
$^3$\mbox{Department of Physics, University of South Africa, Johannesburg 1710, South Africa} \\
$^4$ \mbox{Dubna State University, Dubna, Moscow Region, Russia}\\
 $^5$ \mbox{Moscow Institute of Physics and Technology, Dolgoprudny, 141700, Russia}\\
}
\date{\today}

\begin{abstract}
Recently an irregular easy axis reorientation demonstrating the Kapitza pendulum features were observed in numerical simulations of nanomagnet coupled to the Josephson junction. To explain its origin  we study the magnetization bifurcations and chaos which appear in this system due to interplay of superconductivity and magnetism. The bifurcation structure of the magnetization under the variation of Josephson to magnetic energy ratio as a control parameter demonstrates several precessional motions. They are related to chaotic behavior, bistability, and multiperiodic orbits in the ferromagnetic resonance region.  Effect of external periodic drive on the bifurcation structure is investigated. The results demonstrate high-frequency modes of periodic motion and chaotic response near resonance. Far from the ferromagnetic resonance we observe a quasiperiodic behavior.
\end{abstract}

\pacs{05.70.Ln, 05.30.Rt, 71.10.Pm}

\maketitle
\section{Introduction}  \label{sec_intro}

Spintronics is currently the main contender for next-generation nanoscale devices, aiming for faster processing speeds and lower power consumption~\cite{Hirohata2020}. On the other hand, superconductors stand out as ultra-low energy dissipation systems. Superconductivity thus has the potential to reduce inherent heating effects in spintronic devices. As such, many different approaches have been developed to enhance spintronic effects through the incorporation of superconductivity, and to understand the interactions that arise due to the coexistence of superconducting and magnetic states. Such efforts have spawned the relatively new field of superconductor spintronics~\cite{Golubov2017,Linder2015}.

Molecular nanomagnets~\cite{Rocha,Bogani,Candini2011} are good candidates for qubit realization, due to their long magnetization relaxation time~\cite{Ardavan2007, chiesa2020, Cirillo2019, Roopayan}. Hybrid structures, such as the nanomagnet coupled to Josephson junction (NM-JJ), are also important contenders for the development of spintronic devices~\cite{cai2010interaction,snrk-jetpl_19}. The dynamics of magnetic nanoparticles and that of the JJ are separately governed by nonlinear differential equations. The magnetic nanoparticle can be described by the Landau–Lifshitz-Gilbert equation \cite{Lakshmanan2011}, while, the Josephson junction, can be described by the resistively and capacitively shunted Josephson junction (RCSJ) model~\cite{buc04}.

The coupling in JJ-NM system may be established in different ways, particularly,  through the spin orbit  coupling in $\varphi_{0}-$ junction \cite{buzdin2008direct}. Another type of coupling is realized in NM-JJ,  where the electromagnetic coupling between spin-wave and Josephson phase takes place \cite{weides20060,pfeiffer2008static,hikino2011ferromagnetic,wild2010josephson,kemmler2010magnetic,volkov2009hybridization,mai2011interaction,Nashaat2018}.

The nonlinear dynamics of the JJ is sensitive to the orientation of the magnetization \cite{Buzdin2005,Rabinvich2018,moor2015,minrov2015,silaev2017,bobkova2017,Shukrinov2017,Nashaat2019}, and a rich physics has been predicted due to this type of coupling between the Josephson and magnetic subsystems: for example, supercurrent-induced magnetization dynamics \cite{Golubov2004,Buzdin2005,Linder2014}. In the NM-JJ system, the reversal of the magnetic moment by the supercurrent pulse~\cite{Shukrinov2017}, the appearance of Devil's staircase~\cite{Nashaat2018} and Kapitza pendulum effects~\cite{Shukrinov_epl_2018,snrk-jetpl_19,Kirill_2021arXiv}, have been investigated.

In Ref. \cite{snrk-jetpl_19,Kirill_2021arXiv}, the authors introduced the Kapitza pendulum as a mechanical analog to the NM-JJ system and demonstrated the reorientation of the easy axis of the magnetic moment of the nanomagnet.  In this case, the Josephson to magnetic energy ratio $G$ plays the role of the drive amplitude of the variable force and Josephson frequency $\Omega_J$ plays the role of the drive frequency in the Kapitza problem. The average magnetization component $m_z$ characterizes the changes of the stability position. However, at present, to the best of our knowledge, there is no systematic study of the nonlinear dynamic features in the NM-JJ systems. Therefore, in this article, the dynamical equations which describe the coupling in this system in the framework of the voltage biased Josephson junction is studied. We investigate the magnetization bifurcations and chaos which appear in this system due to interplay of superconductivity and magnetism, and calculate the bifurcation diagrams, Lyapunov exponents and Poincar\'e sections. The several precessional motions related to chaotic behavior, bistability, and multiperiodic orbits in the ferromagnetic resonance region (FMR) are demonstrated. Chaos driven by the external periodic drive is also investigated. An estimation of the model parameters shows that there is a possibility for the experimental observation of the predicted phenomenon.

The plan of the rest of the paper is as follows. In Sec.\ref{NM-JJ}, we describe the model and present an estimation of the model parameters. The dynamics and reorientation features of the nanomagnet coupled to Josephson junction is demonstrated in Sec.\ref{NM_coupled_JJ}. This is followed by a discussion of the  bifurcation diagrams and Poincar\'e sections. In Sec.\ref{rad-effect} we discuss the chaos driven by an external periodic drive. Here, the appearance of the quasiperiodic motion is presented also. Finally, we conclude in Sec.\ref{conclusion}.

\section{Model}  \label{NM-JJ}

\begin{figure}[h!]
	\centering
	\includegraphics[width=0.7\linewidth, angle =0]{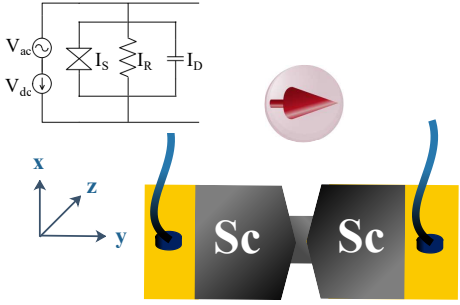}
		\caption{Schematic diagram of the system of JJ-NM with the system geometry. In the equivalent circuit, $V_{dc}$ is the bias voltage, $V_{ac}$ is the ac external drive,  $I_{s}$ is the superconducting current, $I_{R}$ is the resistive current, and $I_{D}$ is the displacement current. }
	\label{JJ-NM}
\end{figure}

We consider short Josephson junction (JJ) with length $l$ coupled to a single-domain nanomagnet with magnetization $\textbf{M}$ and easy axes in $y-$direction. The nanomagnet is located at distance $r_M = a \hat{e}_x$ from the center of the junction as shown in Fig.\ref{JJ-NM}. The magnetic field of the nanomagnet alters the Josephson current, while the magnetic field generated by the Josephson junction acts on the magnetization of the nanomagnet. Thus, an electromagnetic interaction between the JJ and nanomagnet is occurred.

The magnetization dynamics in such system can be described by the Landau–Lifshitz–Gilbert (LLG) equation \cite{landau,gilbert}:

\begin{eqnarray}
	\dfrac{d\textbf{M}}{dt} = -\gamma \textbf{H}_{eff} \times \textbf{M}+\dfrac{\alpha}{M_0}\left(\textbf{M}\times\dfrac{d\textbf{M}}{dt}\right),
	\label{LLG}
\end{eqnarray}
where $\alpha$ is the Gilbert damping parameter, which is a property of the material and lies between 0.0001 and 0.1 for most ferromagnetic materials \cite{Velez,ferona2017}, and $\gamma$ is the gyromagnetic factor. The effective field in LLG equation is given by \cite{buzdin2008direct}:
\begin{eqnarray}
	H_{eff} &=& -\frac{1}{V_{F}} \dfrac{\partial E}{\partial M},
	\label{Heff}
\end{eqnarray}
where $V_{F}$ is the volume of the nanomagnet, and the total energy (E) of the system is the sum of magnetic anisotropy energy ($E_{M}$), Josephson energy ($E_{J}$), and Zeeman energy ($E_{Z}$). The first two terms are given by
\begin{eqnarray}
	E_{M}&=&-\frac{K_{an}  V_{F}}{2} \; \bigg(\dfrac{M_y}{M_0}\bigg)^2,  \\
	E_J &=& \epsilon_J \bigg[1 - \cos\bigg(\frac{2\pi }{\Phi_0}v t +\varphi_m\bigg)\bigg].
\end{eqnarray}

Here  $K_{an}$ is the magnetic anisotropy constant, \mbox{$ M_0=\mid\textbf{M}\mid $} is the saturation magnetization, $\epsilon_{J}=\Phi_{0}I_{c}/2\pi$, $I_{c}$  is critical current of the JJ, $\Phi_{0}$ is the flux quantum, and $v$ is the bias voltage for JJ. The phase shift $\varphi_m $ is induced due to mutual interaction of the nanomagnet and JJ. This shift can be calculated from the vector potential $\textbf{A}_{m}(\textbf{r}, t)$ which takes into account the magnetic field of the nanomagnet created at point $r$ and external magnetic fields if considered (see refs. \cite{cai2010interaction, Kirill_2021arXiv} for detail). According to this, the shift is given by \cite{cai2010interaction}:
\begin{eqnarray}
	\varphi_m &=& - \dfrac{2\pi}{\Phi_0} \int d\textbf{\textit{l}} \cdot \boldmath{\textbf{A}_{m}(\textbf{r},t)} = - k m_z, \nonumber\\
	\textbf{A}_{m}(\textbf{r}, t) &=& \frac{\mu_0}{4 \pi} \dfrac{\textbf{M} \times \textbf{r}}{r^3}, \nonumber \\
	k &=& \dfrac{2\pi}{\Phi_0} \; \dfrac{\mu_{0} M_{0}l}{a \sqrt{a^2+l^2}},
	\label{Phase_dif}
\end{eqnarray}

where the integration goes from one side of the junction to the other side, $\mu_{0}$ is the permeability of free space and  $k$ play the role of the coupling in the proposed system. The last term which contributes to the total energy is generated by the normal current and is given by \cite{cai2010interaction,Kirill_2021arXiv}:
\begin{eqnarray}
	E_z &=& - I_N \int d\textbf{\textit{l}} \cdot \textbf{A}_{m}(\textbf{r}, t).
	\label{jose_en}
\end{eqnarray}

where in the dimensionless form  $I_N =[ V - k \dot{m_z}]$, and $V=v/ I_{c} R$ is the normalized voltage. In our normalization  $V=\Omega_{J}$, $\Omega_{J}=\omega_J/\omega_{c}$, $\omega_J$ is the Josephson frequency $\omega_J=2 \pi v/\Phi_{0} $, $\textbf{m}=\textbf{M}/M_{0}$, $t$ is normalized to $\omega_{c}^{-1}$, $\omega_{c}=2\pi I_{c}R/\Phi_{0}$ is the Josephson characteristic frequency, $R$ is the junction resistance, $\omega_{F}$ is the ferromagnetic resonance frequency, $\Omega_{F}=\omega_{F}/\omega_{c}$, and the effective field $\textbf{h}_{eff}$ is normalized to magnetic anisotropy field. According to this, the LLG reads as

\begin{equation}
	\dfrac{d\mathbf{m}}{dt} = - \frac{\Omega_{F}}{(1+\alpha^{2})}\bigg( \mathbf{m} \times
	\mathbf{h}_{eff} + \alpha \left[ \mathbf{m} \times ( \mathbf{m} \times \mathbf{h}_{eff})\right] \bigg),
	\label{eq5}
\end{equation}
with,
\begin{eqnarray}
&&	h_y=m_y \label{easy}, \;\;  \;\;	h_z=\tilde{h}_{z}-\epsilon k \dot{m_z}  \nonumber \\
&&\text{and}\;\;  \;\; \tilde{h}_{z} = \epsilon [\sin(\Omega_{J} t - k m_{z}) + \Omega_{J}].
	\label{h_eff}
\end{eqnarray}

where $h_y$, $h_z$ are the components of the effective field in the y- and z-direction respectively, $\epsilon = G k$, $G = \epsilon_J / K_{an} V_{F}$ is the Josephson to magnetic energy ratio.

For experimental realization of such a system, we introduce approximate estimations for the model parameters based on Refs.\cite{ Mangin,Cowburn,Yin,Buschow}. We present in Table.\ref{table1} estimations for typical Josephson junctions, in Table.\ref{table2} for typical nanomagnet parameters, and in Table.\ref{table3} for model parameters. The value of $k$ depends on the distance of the nanomagnet from the JJ and the length of the junction (here, for estimation we consider $a=250 \mu m$).  Experimental results give the estimation for the ferromagnetic resonance frequency of nanomagnets within the range of $\sim5-10  GHz$ \cite{Nekrashevich,Kachkachi2018}. In the voltage-biased Josephson junction, one can tune the Josephson frequency in a wide region around the FMR.
\begin{table}[ht]
	\centering
	\caption{Typical Josephson junctions parameters}
	\label{table1}
	\begin{tabular}{|c|c|c|}
		\hline
		Parameter
		& $Al/Al_{2}O_{3} /Al$
		&  $Nb/Al_{2}O_{3} /Nb$ \\
		\hline
		$l$
		&141 $n m$
		& 20 $n m$ \\
		\hline
		$I_{c}$
		& 20 $nA$
		& 6 $mA$ \\
		\hline
		$\epsilon_{J}$
		&6.58 $\times$ 10$^{-24}$ $J$  & 2.19 $\times$ 10$^{-18}$ $J$   \\
		\hline
		$R$
		&10 $k\Omega$
		& 0.003$\Omega$
		\\
		\hline
	$\omega_{c}$ 	& $\sim$ 600 $GHz$  &   $\sim$ 50 $GHz$  \\
	\hline
	\end{tabular}
\end{table}

\begin{table}[ht]
	\centering
	\caption{Parameters}
	\label{table2}
	\begin{tabular}{|c|c|}
		\hline
		Parameter
		& Material \\
		\hline
		\shortstack{$M_{0}$ \\ \\ \ }
		&\shortstack{ 907 $kA/m$ ($SmCo_{5}$),\\  1950 $kA/m$   ($Fe_{65}Co_{35}$)} \\
		\hline
		\shortstack{$K_{an}$\\ \\ \ }
		&\shortstack{  17000 $kJ/m^{3}$ ($SmCo_{5}$),\\  20 $kJ/m^{3}$  ($Fe_{65}Co_{35}$)}\\
		\hline
		 $v$
		&  $\sim$1.979 $m^{3}$ $\times$ 10$^{-23}$ \\
		\hline
	\end{tabular}
\end{table}

\begin{table}[h!]
	\centering
	\caption{Model parameters}
	\label{table3}
	\begin{tabular}{|c|c|c|}
		\hline
		Parameter
		&$Al/Al_{2}O_{3} /Al$  &  $Nb/Al_{2}O_{3} /Nb$ \\
		\hline
		$k$	& 0.01 &   0.01  \\
		\hline
		\shortstack{ $G$\\ \\   \\  \\\ }	& 	\shortstack{0.0001\\ \\  \\  \\  \\\ } & \shortstack{5.5, \\ $10\pi$( $v \sim$ 120 $nm^{3}$) \\ and $k_{an}=10 kJ/m^{3}$} \\
		\hline
	\end{tabular}
\end{table}

The results presented in the paper have been obtained using different numerical methods. In particular, we solve Eq.(\ref{LLG}) numerically using implicit Gauss-Legendre method \cite{shuk-gl4-2019} to calculate the dynamics of the system. To characterize different kind of motions the bifurcation diagram, the Poincar\'e sections and the largest Lyapunov exponent (\textit{LLE}) have been calculated. In this case, we solve the system of equations (\ref{LLG}) in the fixed time interval which is multiplied to the drive period ($ 2 \pi / \Omega_J$) with the time step equals to $\sim 10^{-4}$. The number of time steps in our time domain is equal $10^8$. Then, we save the values of the components $m_x, m_y,$ and $m_z$ at the end of the time interval, those points also create the Poincar\'e section for trajectories in phase space. To find the \textit{LLE} as a function of $G$, we calculate the magnetic moment dynamics with the initial conditions $m=(0,1,0)$, then randomly shift them ($\delta \approx 10^{-5}$) from the reference one. Then, we calculate the \textit{LLE} from $m_x, m_y, m_z$ and $m_x + \delta, m_y + \delta, m_z + \delta$ and average it over time. Of course, different initial conditions can lead to slightly different pictures, but the qualitative picture remains the same. The \textit{LLE} is used to determine the sensitivity of the system to the initial conditions. The positive \textit{LLE} is one of the signs of chaos in the system, this means that two phase space trajectories with a small difference in initial conditions will rapidly diverge, and then have totally different futures. The negative value of \textit{LLE} indicates that the system approaches a fixed point (here the fixed point is $(0,0,1)$ for which $<m_z(t)>=1$). The zero value of \textit{LLE} shows that the system is periodic or quasiperiodic.

We consider the ferromagnetic resonance frequency $\Omega_F = 1$, the coupling constant between the JJ and the nanomagnet $k = 0.05$, and the Gilbert damping parameter $\alpha = 0.1$. We have chosen the Josephson to magnetic energy ratio $G$ and the Josephson frequency $\Omega_J$ as control parameters, which represents an experimentally reasonable choice. All our calculations start with minimum value of \textbf{$G=0.01\pi$}.

\section{Irregular reorientation behavior, bifurcations and chaos}  \label{NM_coupled_JJ}

In the proposed model, the magnetic field of the total tunneling current (both the superconducting and quasiparticle) have been taken into account. This leads to the two different reorientation mechanisms of the nanomagnet easy axis. One mechanism is related to the magnetic field, created by the quasiparticle current flowing through the JJ. The other one is related to the oscillating magnetic field generated by the superconducting current.  The second one is a manifestation of Kapitza pendulum-like feature which was observed in the magnetization dynamics of the nanomagnet \cite{snrk-jetpl_19,Kirill_2021arXiv} and $\varphi_0$-junction \cite{Shukrinov_epl_2018}.

\begin{figure}[h!]
	\centering
	\includegraphics[width=0.8\linewidth, angle =0]{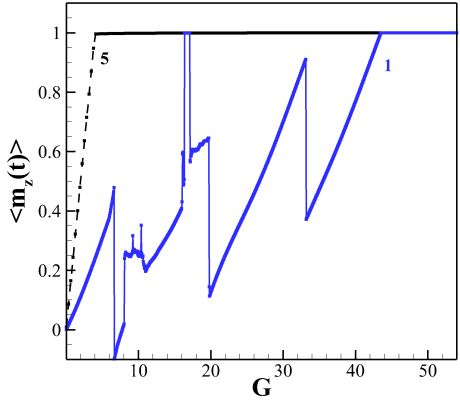}
	\caption{The average magnetization component $<m_z(t)>$ as a function of $G$, demonstrating the Kapitza pendulum-like features in NM-JJ system \cite{snrk-jetpl_19,Kirill_2021arXiv}. The blue solid line indicate the results calculated at $\Omega_J=1$. The black dashed line at $\Omega_J=5$.  }
	\label{kapitza}
\end{figure}

Figure \ref{kapitza} shows the average magnetic moment component $m_z$ as a function of the Josephson to magnetic energy ratio $G$. One can see the smooth change of $<m_z(t)>$ from zero to one as a function of $G$ at $\Omega_{J}=5$. The stabilization of the magnetic moment components dynamics occurs at $M = (0, 0, 1)$, when $G$ exceeds a certain reorientation value, which indicates a complete reorientation of the magnetic moment. Notice that for $\Omega_{J}\simeq\Omega_{F}$ the fluctuations of $<m_z(t)>$ appear before the complete reorientation. To understand the origin of the fluctuations at the FMR, we investigate the dynamics of the effective field and the transformations of the magnetic moment dynamics ifor two cases, one at the FMR, the other far away.

%

\begin{figure}[h!]
    \centering
	\begin{minipage}{0.48\textwidth}	
		\centering
		\includegraphics[width=0.48\linewidth, angle =0]{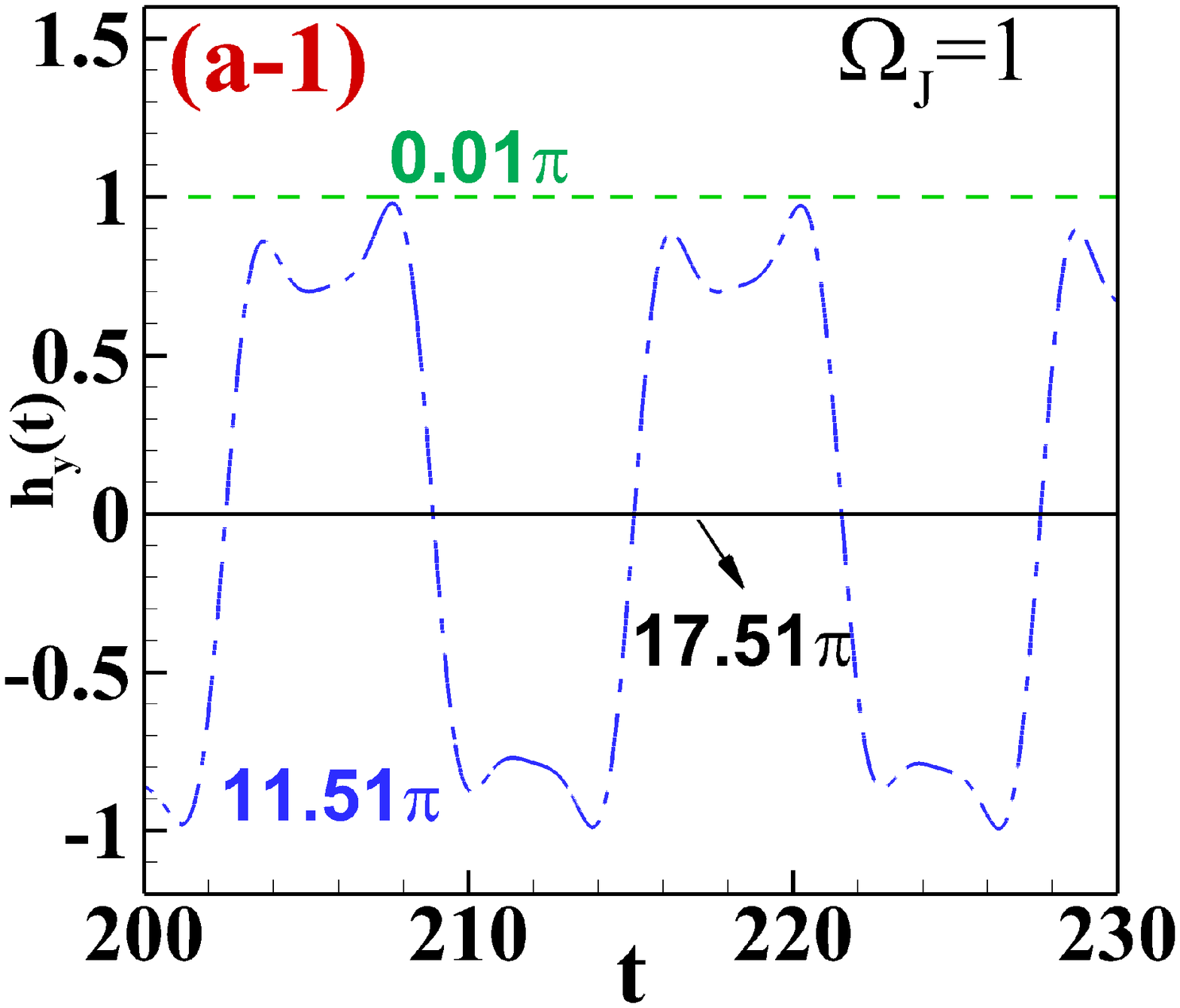}
		\includegraphics[width=0.48\linewidth, angle =0]{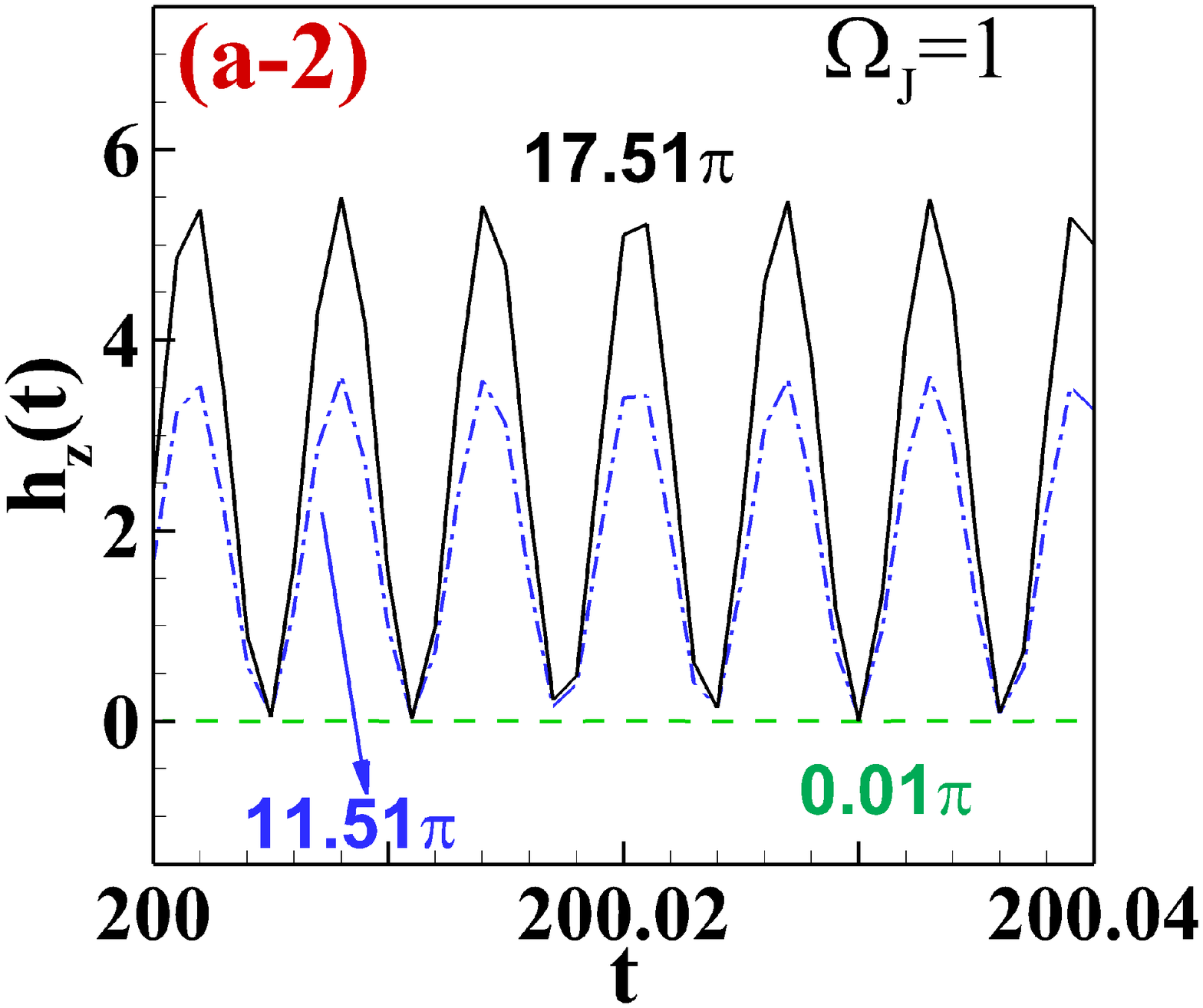}
		\includegraphics[width=0.48\linewidth, angle =0]{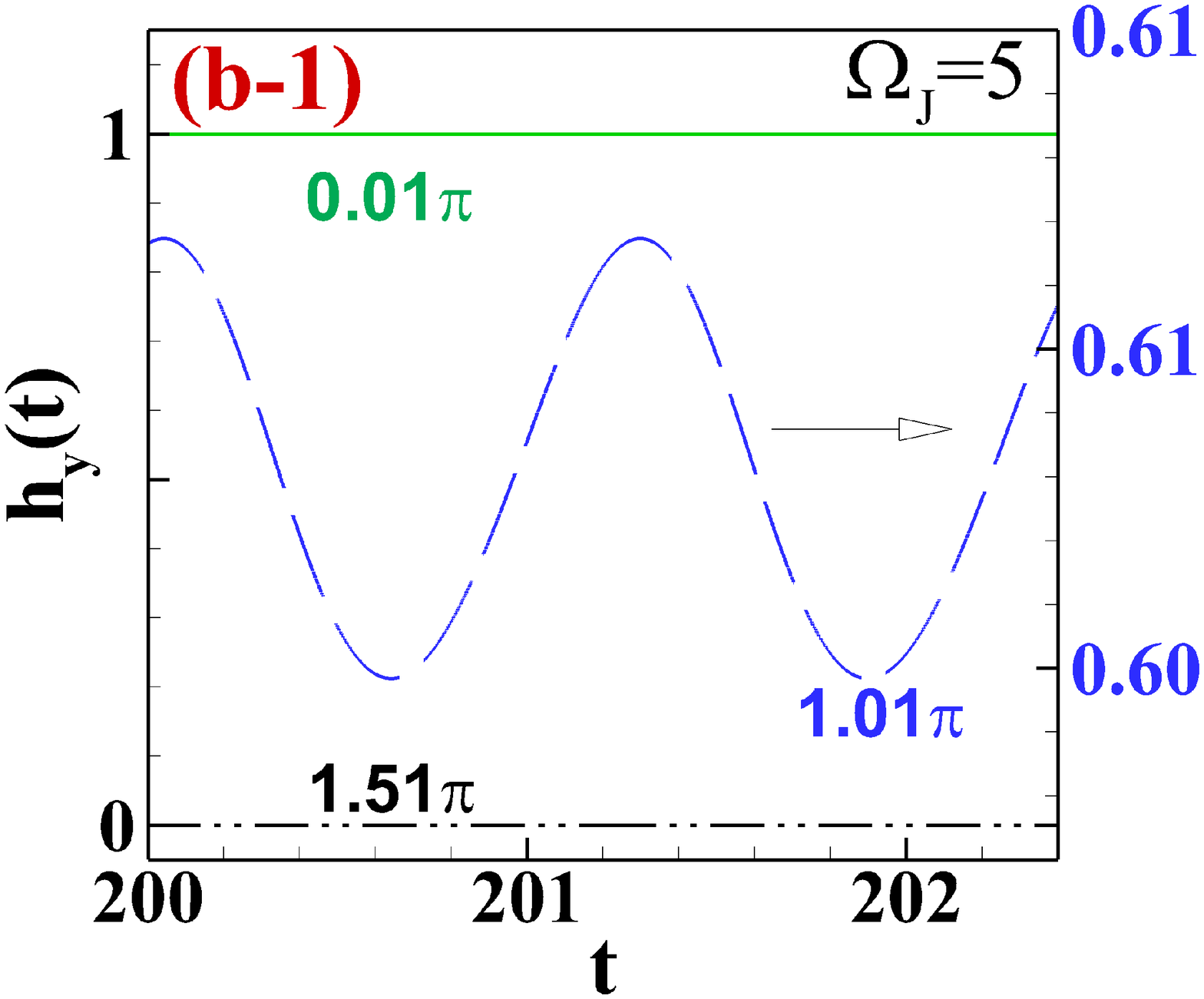}
		\includegraphics[width=0.48\linewidth, angle =0]{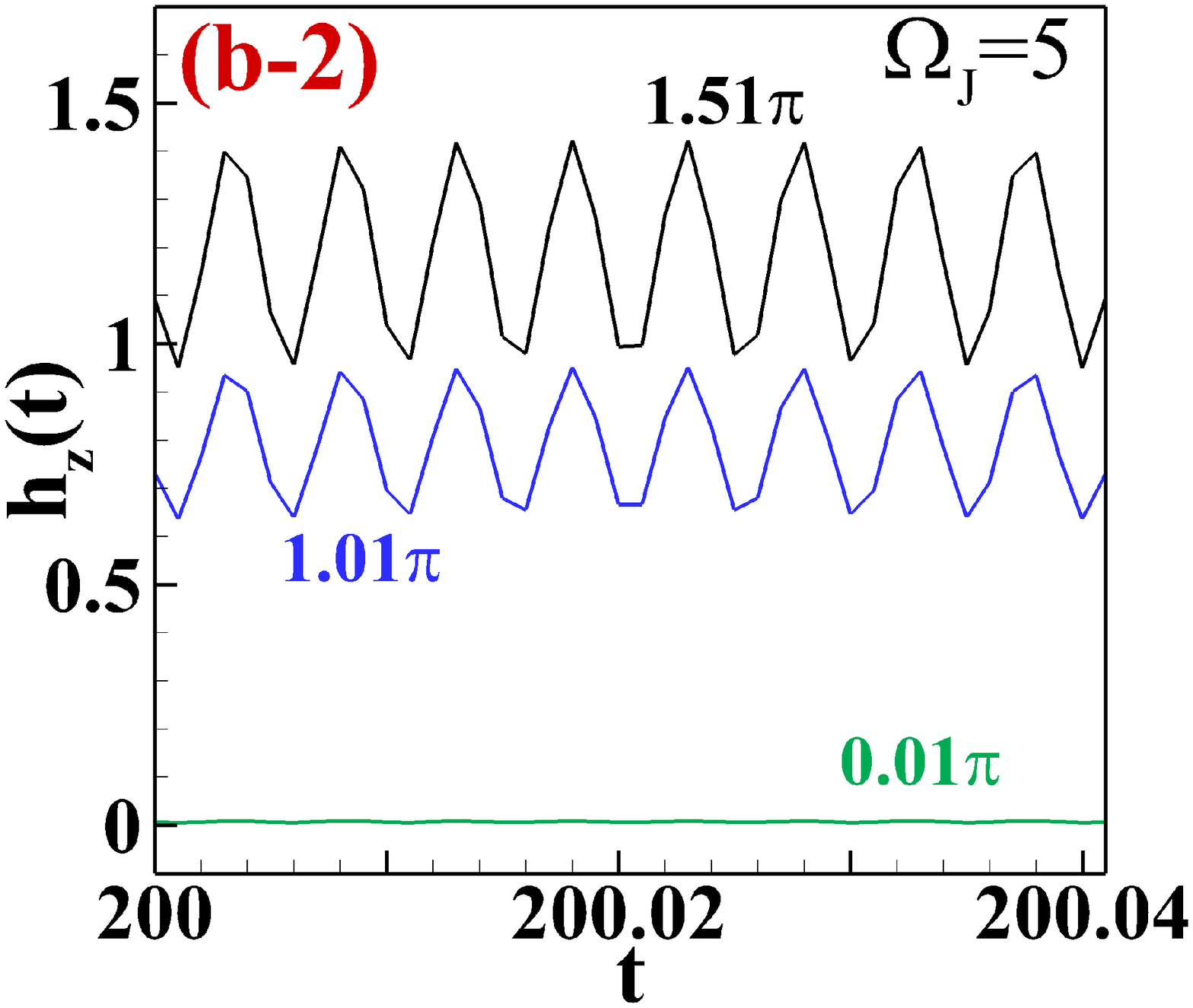}
		\caption{The temporal dependence of effective field components as a function of $G$: (a-1) $h_y(t)$ and (a-2) $h_z(t)$ at $\Omega_J = 1$, (b-1) $h_y(t)$ and (b-2) $h_z(t)$ at $\Omega_J = 5$. The numbers indicate the value of $G$.}
		\label{eff_wj}
	\end{minipage}
\end{figure}

The dynamics of the effective field components $h_{y}(t)$ and $h_{z}(t)$ as functions of $G$ at $\Omega_{J}=1$ and $\Omega_{J}=5$ are demonstrated in Fig.~\ref{eff_wj}. As we see, there is no temporal dependence of $h_{y}(t)$ and $h_{z}(t)$ at a small $G<<1$ (the curve with $G=0.01\pi$) since the Josephson energy is too small in compare to anisotropy energy and its magnetic field does not affect the nanomagnet. The temporal dependence of $h_y(t)$ at $G>1$ demonstrates irregular oscillations with different amplitudes at $\Omega_{J}\simeq\Omega_{F}$ (see Fig.\ref{eff_wj}(a-i)) and regular oscillations at $\Omega_J> \Omega_F$ (see Fig.\ref{eff_wj}(b-i)). On the other hand, the temporal dependence of $h_{z}(t)$ shows a periodic structure (see Fig.\ref{eff_wj}(a-ii) and Fig.\ref{eff_wj}(b-ii)) with amplitude which increases with the increasing in $G$ at fixed $\Omega_J$. The stabilization of the magnetic moment dynamics occurs at $M = (0, 0, 1)$, when $G$ exceeds a certain reorientation value, which indicates a complete reorientation of the magnetic moment. The value of $G$ at which the complete reorientation occurs decreases with increasing $\Omega_{J}$.

\begin{figure}[h!]
	\centering
	\includegraphics[width=0.6\linewidth, angle =0]{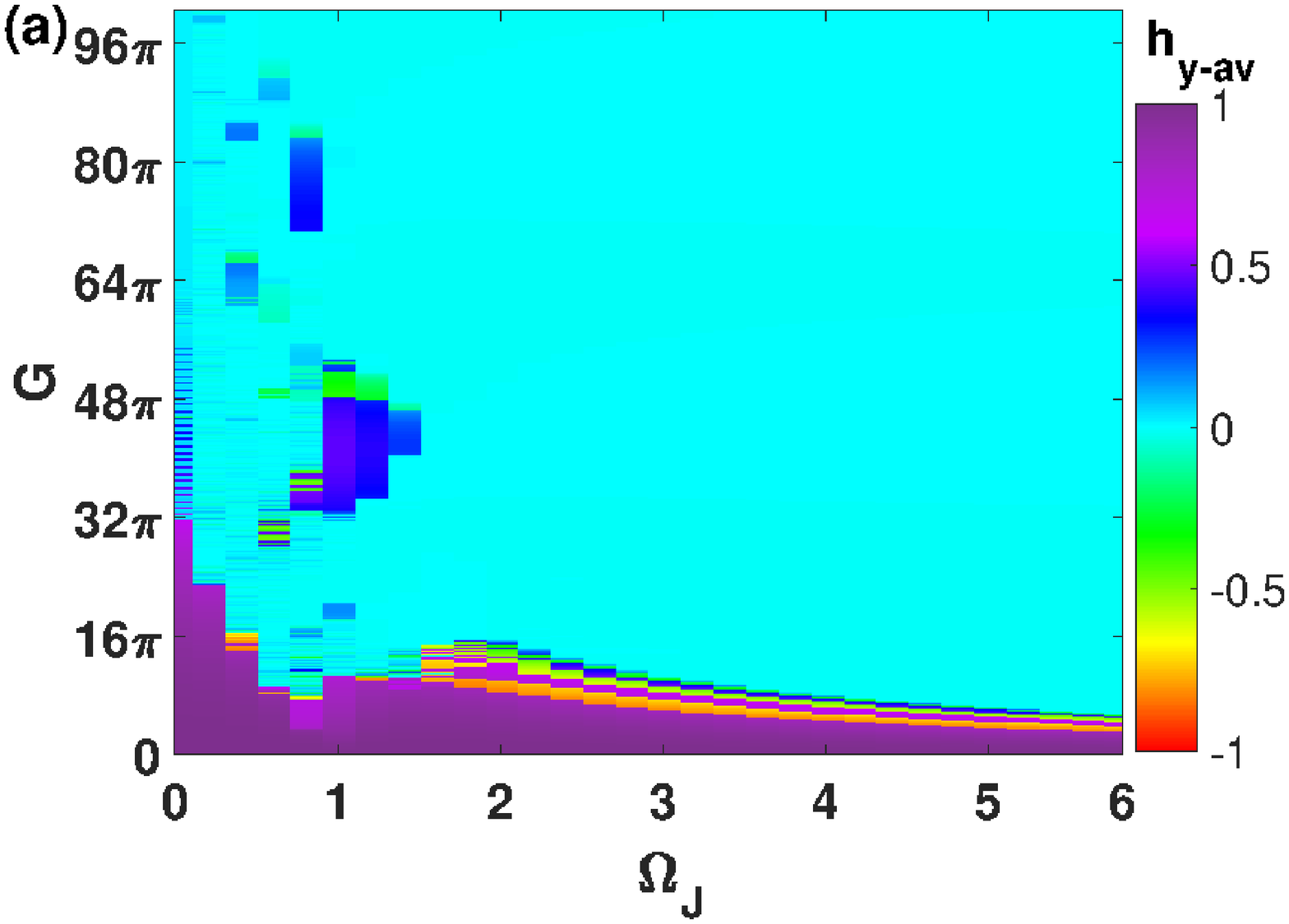}
	\includegraphics[width=0.6\linewidth, angle =0]{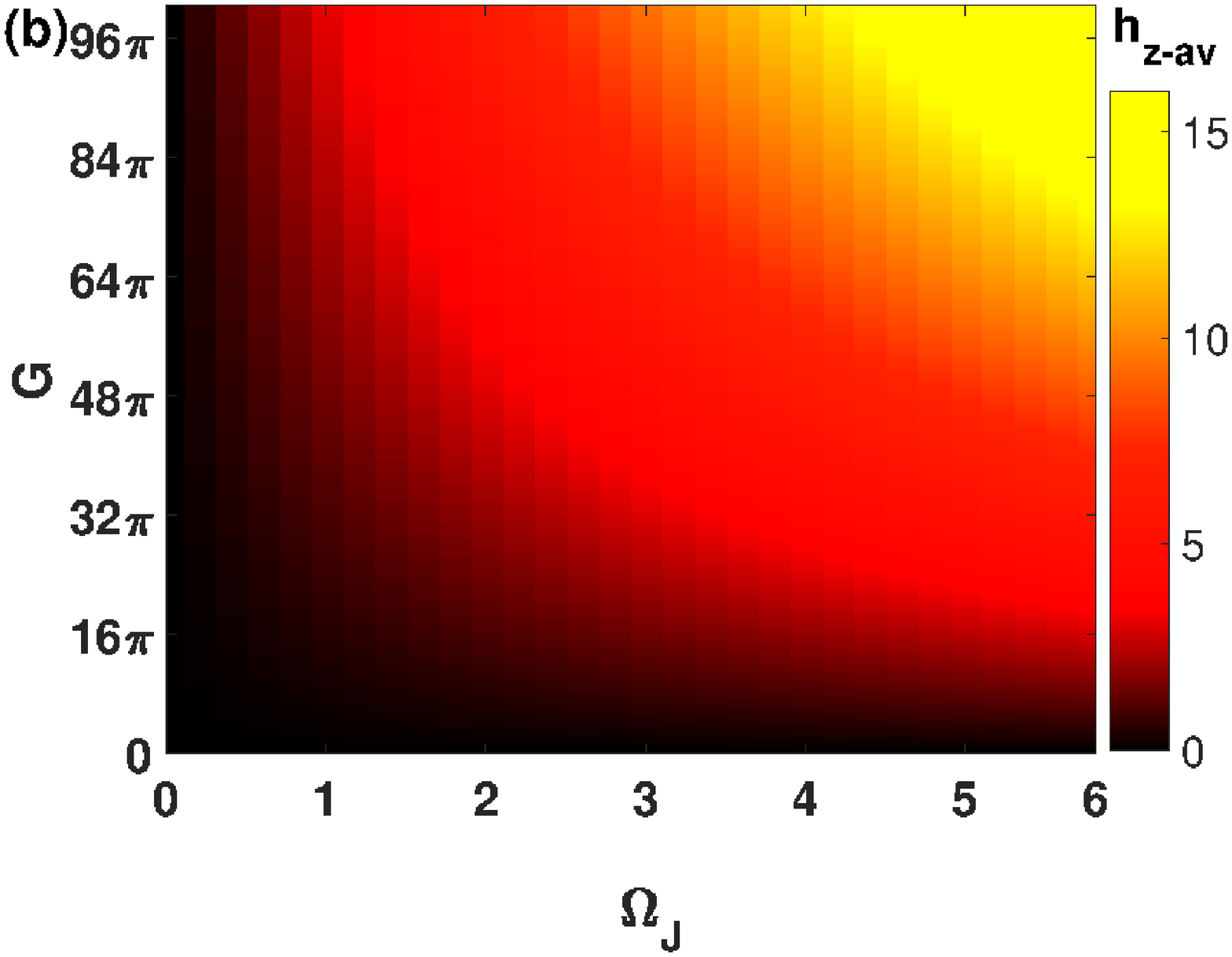}
	\includegraphics[width=0.6\linewidth, angle =0]{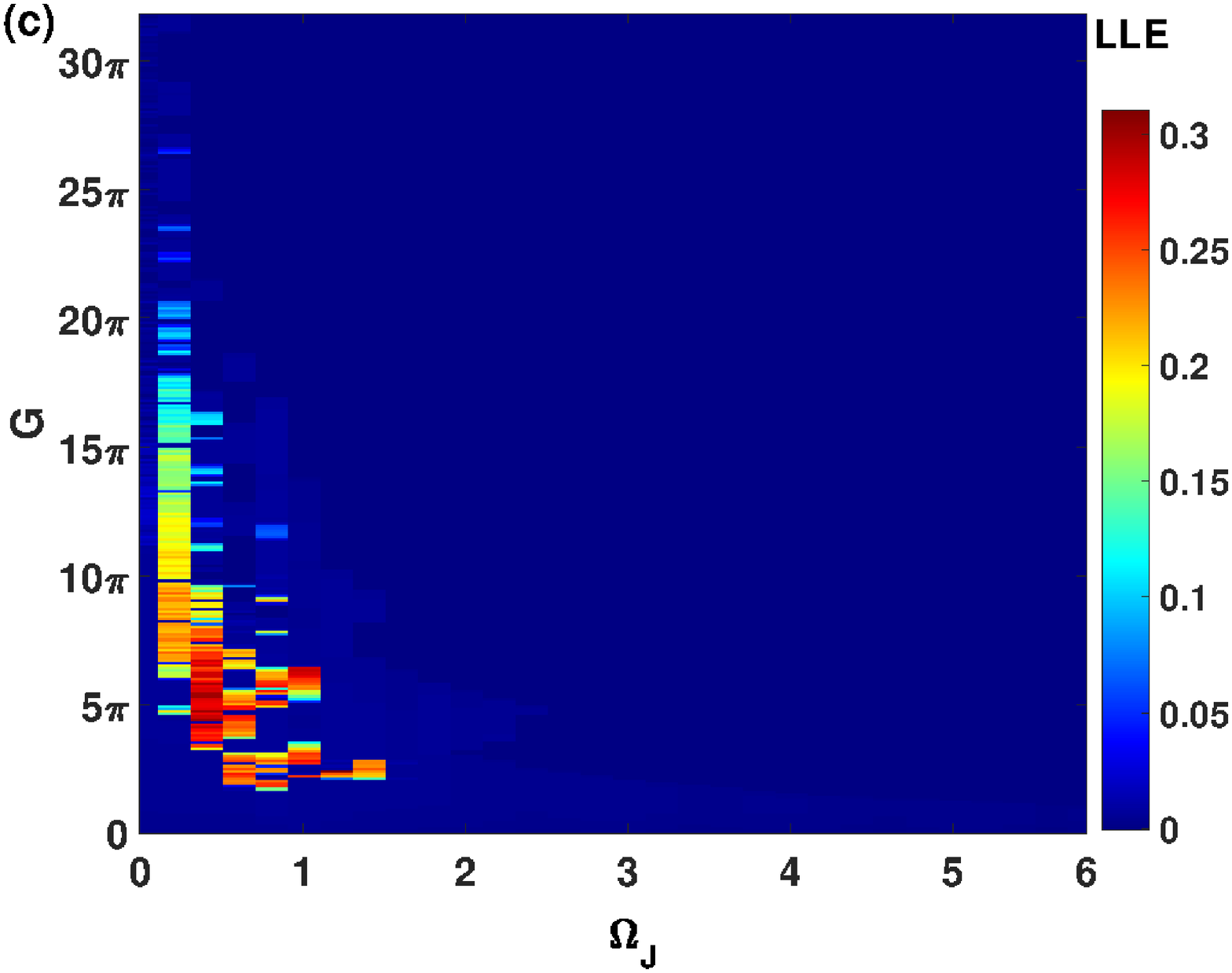} 	
	\caption{ (a) The average value of $h_{y-av}$, (b) the average value of $h_{z-av}$  and (c) the largest Lyapunov exponent as functions of $G$ and $\Omega_J$. }
	
	\label{heff_wj_G}
\end{figure}

The dynamical behavior of $h_{y}(t)$ and $h_{z}(t)$ reflects on the average value of the effective field components. Therefore, we investigate $h_{y-av}$ and $h_{z-av}$ as functions of $G$ and $\Omega_J$ and create 2-D maps, which are demonstrated in Fig.~\ref{heff_wj_G}(a and b). Fig.~\ref{heff_wj_G}(a) shows that the average of $h_{y}$ has a non-zero values only at $G<20\pi$ and around the FMR condition ($\Omega_{J}\simeq\Omega_{F}$), while the average of $h_{z}$ smoothly increasing with the increasing in $G$ and $\Omega_J$ (see Fig.\ref{heff_wj_G}(b)). We note that the condition $h_{y-av}=0$ indicates the complete reorientation of the magnetic moment, while the negative values of $h_{y-av}$ indicates the reversal of the easy axis \cite{cai2010interaction,snrk-jetpl_19}. The reorientation features at $\Omega_{J}>>\Omega_{F}$ have been investigated in Refs. \cite{snrk-jetpl_19, Kirill_2021arXiv}. Here, we investigate the regions of the non-zero values of $h_{y-av}$ which appear at small $\Omega_{J}$ ($\Omega_{J}<2$). The system in this region is influenced by the irregular oscillations of $h_{y}(t)$, which can be a cause of a chaotic dynamic of the nanomagnet. So, we calculated the \textit{LLE} as functions of $G$ and $\Omega_{J}$.  The results of our calculation is presented in Fig.\ref{heff_wj_G}(c). The \textit{LLE} shows a non zero value at $\Omega_{J}\simeq\Omega_{F}$. Therefore, the system may demonstrates a chaotic response in this interval of frequencies.

\subsection{Dynamical effects at $\Omega_{J}=1$}

To confirm the chaotic nature of the magnetic moment dynamics we calculate the bifurcation diagrams. The bifurcation diagrams reveal structural changes of the motion in the parameter space \cite{ferona2017}.

\begin{figure}[h!]
	\centering
	\includegraphics[width=\linewidth, angle =0]{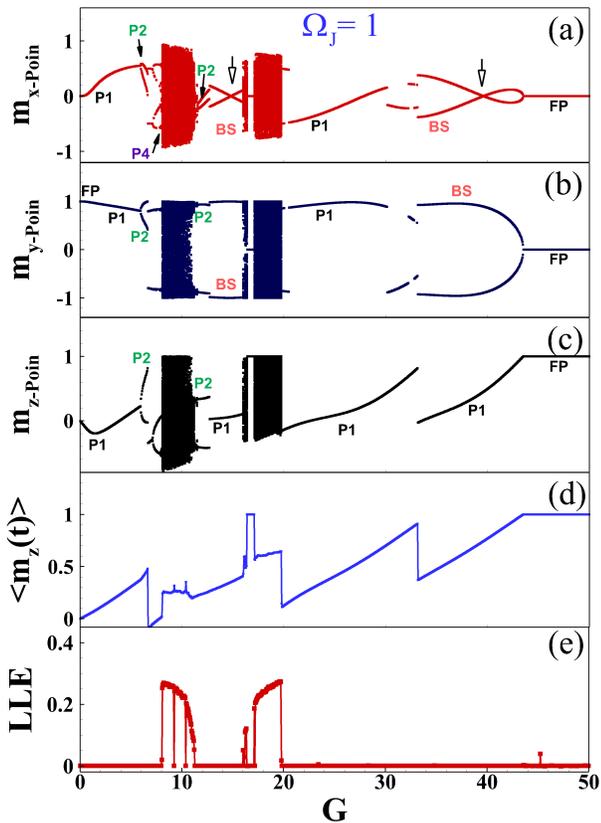}
	\caption{Bifurcation diagram of the magnetization components (a) $m_x$, (b) $m_y$, (c) $m_z$, (d) the average value of $m_z-$component and (e) the largest Lyapunov exponent as a function of $G$ at $\Omega_J = 1$. }
	\label{bifurc_wj_1}
\end{figure}

Figs.\ref{bifurc_wj_1} (a-c) show the bifurcation trees of the magnetization as a functions of $G$ ( at the FMR $\Omega_J = \Omega_F$) before the complete reorientation of the easy axis. The bifurcation tree starts from the fixed point (FP) ($m_y=1$) and demonstrates a period-1 (P1) motion ($0$-branch on the bifurcation tree) at $G < 6$. The first period doubling (or the $1^{st}$ branch) occurs at $G = 6$ and the system starts to perform period-2 (P2) motion up to $G < 7.9$. Second period doubling is observed at $G = 7.9$ (period-4 (P4) motion). Then, chaotic bands can be observed at the intervals $[8,11.5]$ and $[16,20]$, with the periodic motion in between. The system demonstrates a bistable state (BS), two points with the same value, but different signs, for $m_x$ and $m_y$ within the intervals $[12.7,16]$ and $[30,43.5]$, while $m_z$ shows the regular P1 motion. In addition to this, two folding bifurcation are revealed (indicated by the hollow arrows in Fig\ref{bifurc_wj_1}(a)). Finally, the system approaches the stable FP at $G \geq 43.5$, corresponding to the complete reorientation of the easy axis $m_{z}=1$.

The system transition from one kind of motion to another is accompanied by abrupt changes in the average magnetization components. We demonstrate such changes in Fig.\ref{bifurc_wj_1}(d), where the irregular reorientation behavior of $<m_{z}(t)>$ appears before the complete reorientation. The \textit{LLE} calculation confirms the chaotic behavior of the magnetization (see Fig.\ref{bifurc_wj_1}(e)). The positive values intervals of \textit{LLE} coincide with the chaotic bands observed in the bifurcation diagrams.


\begin{figure}[h!]
	\begin{minipage}{\linewidth}	
		\centering
		\includegraphics[width=0.5\linewidth, angle =0]{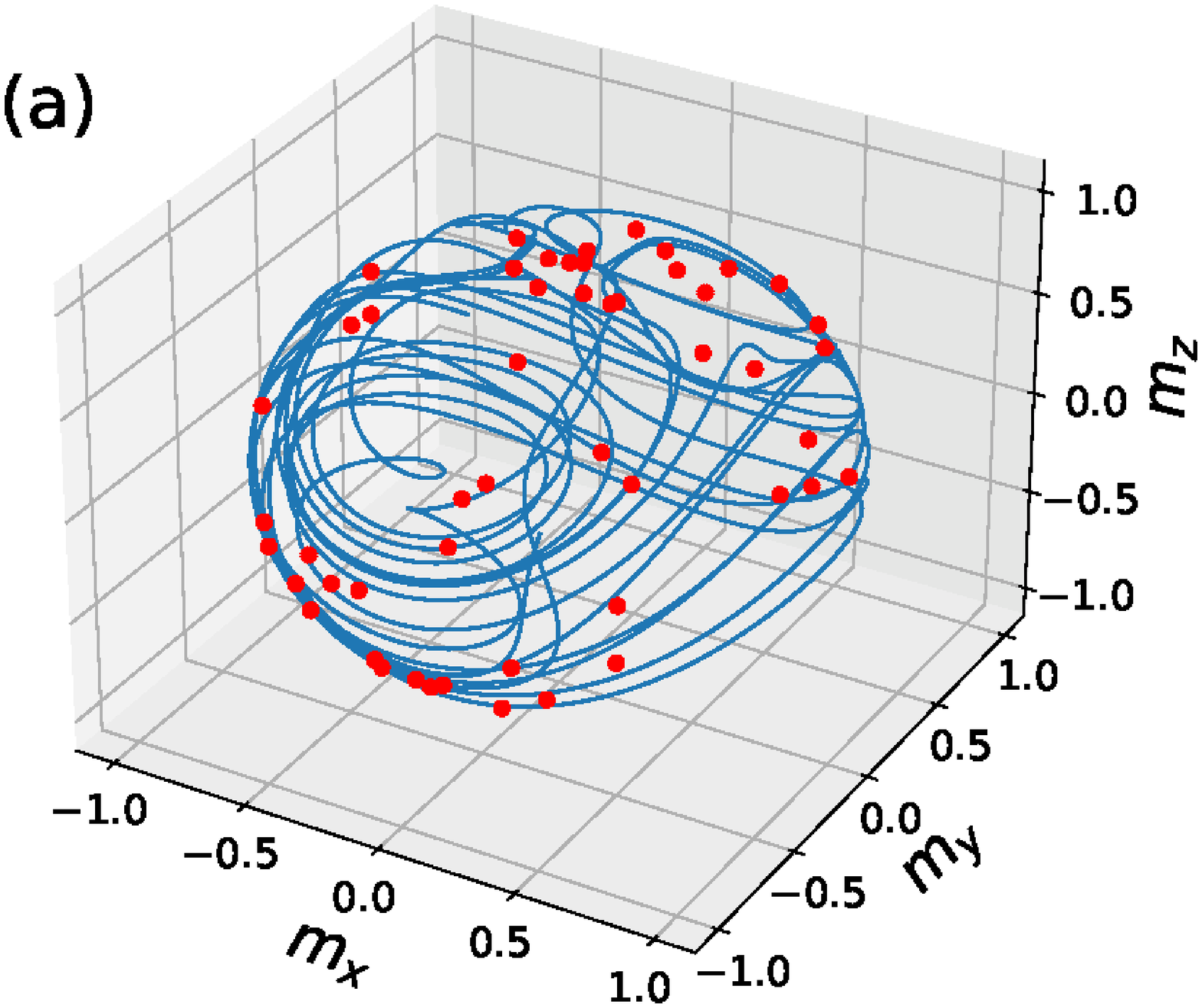}
		\includegraphics[width=\linewidth, angle =0]{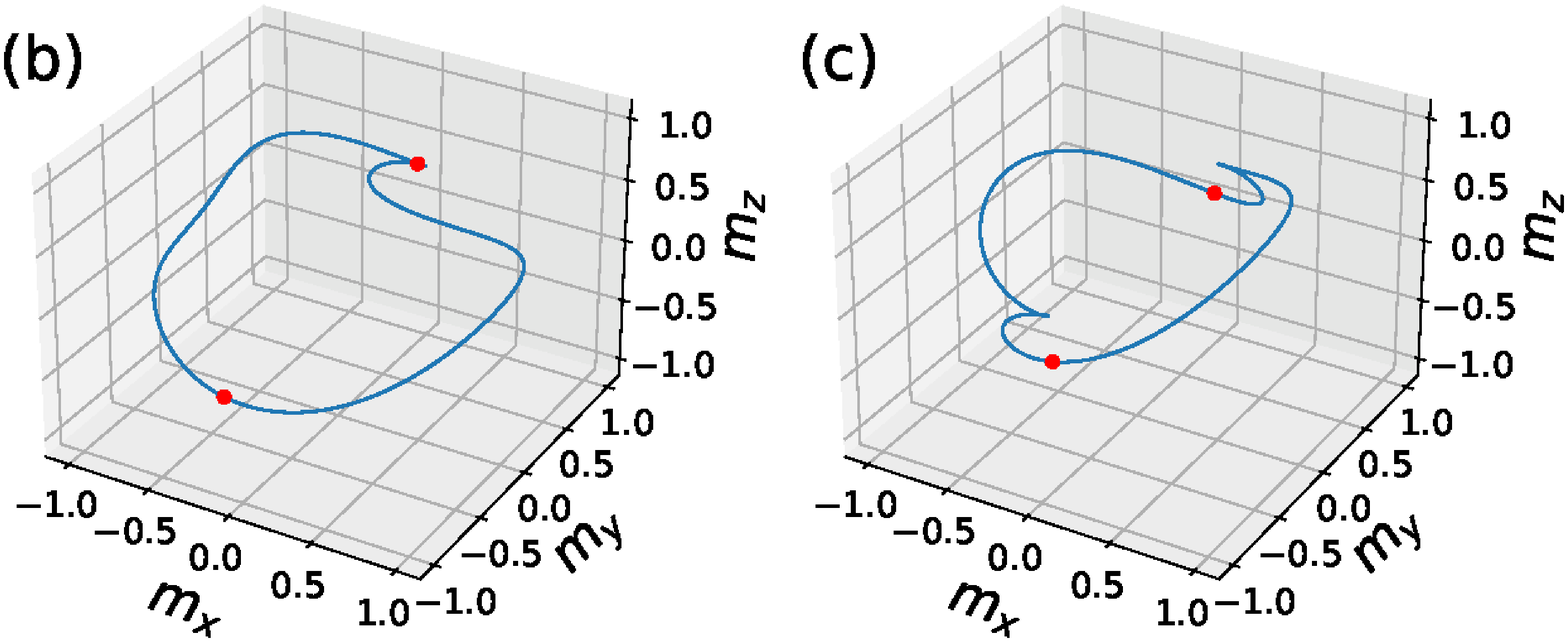}
		\caption{Orbits of motion (blue curve) of the system and corresponding Poincar\'e section (red dots). (a) at $G=9$ with chaos, (b) at $G = 12$ with P2 motion, (c) at $G = 14$ with bistable states in $m_x$ and $m_y$, and P1 motion in $m_z$. All panels are done at $\Omega_J = 1$.}
		\label{attrac_wj_1_g}
	\end{minipage}
\end{figure}

To support our conclusions, concerning the chaotic behavior, we also calculate the Poincar\'e sections along with the orbits of motion at specific values of $G$ and results presented in Fig. \ref{attrac_wj_1_g}. The Poincar\'e sections with one point only indicates that the magnetization exhibits P1 motion, two points -- P2 motion, and so on. The magnetization dynamics shows the dense and random distribution of the trajectories and the Poincar\'e section points  at $G=9$ (see Fig. \ref{attrac_wj_1_g}(a)). This along with the positive \textit{LLE} and dense distribution of points on the bifurcation diagram  confirm the chaotic nature of those states. Fig. \ref{attrac_wj_1_g}(b) demonstrates the trajectories and the Poincar\'e section points of the P2 motion at $[11.5, 12.7]$. However, two points in the Poincar\'e section can also indicate BS, as it is presented in Fig. \ref{attrac_wj_1_g}(c), where $m_x$ and $m_y$ show BS and P1 motion in $m_z$ at $[12.7, 16]$ in the bifurcation diagram. In BS the trajectory reaches a limit cycle near the $\pm y(x)-$axis depending on the initial condition.

\subsection{Dynamical effects at $\Omega_{J}=1.5$}

\begin{figure}[h!]
	\centering
	\includegraphics[width=\linewidth,  angle =0]{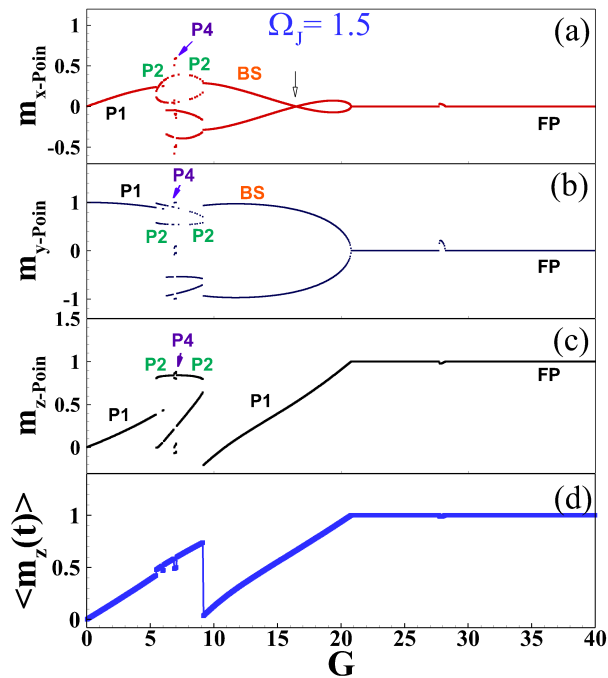}
	\caption{Bifurcation diagram of the magnetization components (a) $m_x$, (b) $m_y$, (c) $m_z$ and (d) the average value of $m_z-$component as functions of $G$ at $\Omega_J = 1.5$.}
	\label{bifurc_wj_1_5}
\end{figure}

Next, we investigate the transformation of the magnetic moment dynamics from the case $\Omega_{J}=\Omega_{F}=1$ to $\Omega_{J}\simeq\Omega_{F}$. Figs.\ref{bifurc_wj_1_5} (a-c) show the bifurcation trees of the magnetization as a functions of $G$ at $\Omega_J = 1.5$. In this case a simpler bifurcation structure is observed for precessional motion in compared to $\Omega_J = 1$. The bifurcation trees at $\Omega_J = 1.5$ demonstrate only the periodic motions of different order, BS and FP and the \textit{LLE} equals zero within the whole calculation range. The average $<m_z(t)>$ as a function of $G$ presented in Fig.\ref{bifurc_wj_1_5}(d) reflects the transition of the system from one kind of motion to another one.

\begin{figure}[h!]
	\begin{minipage}{\linewidth}	
		\centering
		\includegraphics[width=0.7\linewidth, angle =0]{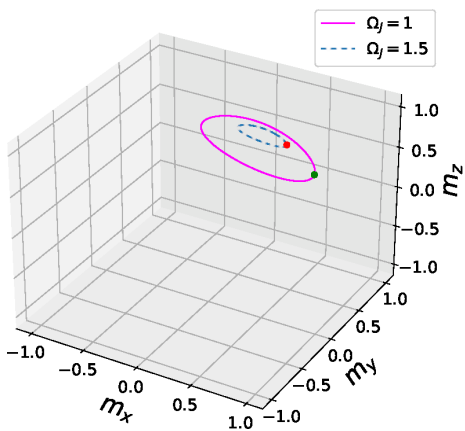}
		\caption{Orbits of the motion of the magnetization (dashed line for $\Omega_J=1.5$, solid line for $\Omega_J=1$) and corresponding Poincaré sections.}
		\label{attrac_wj_1_and_1_5}
	\end{minipage}
\end{figure}

Figure \ref{attrac_wj_1_and_1_5} demonstrates a shrinking of orbits of motion under the increase of $\Omega_J$. The increases of the driving frequency $\Omega_J$ reduces also the value of $G$ at which the complete reorientation occurs (see dashed line in Fig. \ref{kapitza}) \cite{Kirill_2021arXiv,snrk-jetpl_19}. So, only the P1 motion before the stable FP can be observed at $\Omega_J=5$.

In the considered system, the precession of the magnetization is driven by Josephson oscillations. So, the increase in $\Omega_J$ forces the magnetization to follow the Josephson oscillations, and only the periodic motion can be observed. Therefore, it is the reason why we do not observe any chaotic behavior of the magnetization at $\Omega_J \geq 1.5$.

\section{Chaos driven by external periodic drive}  \label{rad-effect}

In this section, we investigate the effect of the external periodic drive (PD) on the magnetization dynamics. In this case the total bias voltage for JJ consists of the dc and ac parts $V_{total} = V + A \cos(\Omega_r t)$, where $A$ is the amplitude of the external radiation normalized to $\hbar \omega_c / 2 e$ and $\Omega_r$ is the frequency of the ac voltage normalized to $\omega_c$. Therefore, the total tunneling current in the JJ is calculated in the framework of RCSJ-model:
\begin{eqnarray}
I(t) &=& \sin\left(\Omega_J t - k m_{z}+\frac{A}{\Omega_r} \sin(\Omega_r t)\right) + \Omega_J + A \cos(\Omega_r t)\nonumber \\&-& k \dfrac{dm_{z}}{d t} - \beta_{c} A \Omega_r \sin(\Omega_r t).
\label{curr_rad}
\end{eqnarray}
The effective field components remain the same as in Eq.(\ref{h_eff}) except $\tilde{h}_{z}$ which is given by:
\begin{eqnarray}
\tilde{h}_{z} &=&   \epsilon \bigg\{\sin\left(\Omega_J t - k m_{z} + \frac{A}{\Omega_r} \sin(\Omega_r t)\right) + \Omega_J \nonumber \\ &+& A \cos(\Omega_r t) - \beta_{c} A \Omega_r \sin(\Omega_r t) \bigg\},
\label{hz_rad}
\end{eqnarray}
where $\beta_c$ is the McCumber’s parameter and the higher-order term $(- \beta_{c} k \ddot{m_z})$ is neglected. In this case, the nanomagnet effective field includes two oscillatory terms. One is the oscillating magnetic field, generated by the superconducting current, with the amplitude proportional to $G$ and with the frequency of Josephson oscillations. The other one is the PD term. The main oscillatory mechanism is determined by the amplitudes of those terms.

\subsection{Bifurcation structure as a function of $G$}

\begin{figure}[h!]
	\centering
	\includegraphics[width=\linewidth, angle =0]{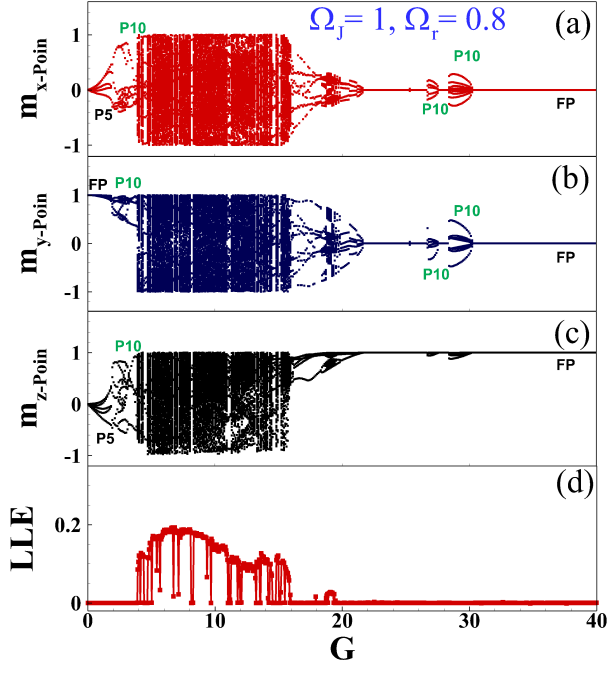}
	\caption{Bifurcation diagrams of the magnetization components (a) $m_x$, (b) $m_y$, (c) $m_z$ and (d) the largest Lyapunov exponent as functions of $G$ at $\Omega_J = 1$ under external periodic drive with $\Omega_{r} = 0.8$ and $A =1$.}
	\label{bifurc_wj_1_A}
\end{figure}

First, we investigate the effect of $G$ on the bifurcation structure under external PD. Figure \ref{bifurc_wj_1_A} shows the bifurcation diagram of the magnetization dynamics and \textit{LLE} at $\Omega_J = 1$, $\Omega_{r}=0.8$ and $A=1$. The bifurcation trees, in this case, starts from the P5 motion as a result of the influence of the external periodic drive (see Fig.\ref{bifurc_wj_1_A}(a)). The magnetization dynamics undergoes the first period doubling at $[2.4, 3.3]$ and after that the system returns back to P5 motion at $[3.3, 3.8]$.  Then, the chaotic band can be observed at the interval $[3.8, 16]$, where the \textit{LLE} values at $[3.9, 16]$ are on the order of $10^{-1}$ (see Fig.\ref{bifurc_wj_1_A}(d)). Inside this chaotic band, there are very narrow windows of periodic motion at several values of $G$, where the P10 motion is observed. Those windows can be distinguished on the \textit{LLE} as a corresponding dips near zero within $[10, 15]$.

A small region of periodic motion with the high order modes is observed at $[18.4, 19.4]$. It is hard to recognize the oscillation modes in this crowded region due to the dense Poincar\'e section points within a small portion of the phase space. However, the \textit{LLE} confirms the periodic nature of motion in this interval since the largest value is on the order of $10^{-2}$ (see Fig.\ref{bifurc_wj_1_A}(d)), which is still small compared to the values of \textit{LLE} where the chaotic behavior observed. The P5 motion appears at $G \geq 19.4$ and the magnetization dynamics approaches FP at $G \geq 21.7$ which is confirmed by the negative value of \textit{LLE} in this region (see Fig.\ref{bifurc_wj_1_A}(d)). However, the system does not settle and shows the P10 motion at the intervals $[25.3, 25.5]$, $[26.6, 27.6]$, and $[28.4, 30.5]$. After that the trajectory finds a stable FP corresponding to a complete reorientation of the magnetization direction ($<m_z(t)> = 1$) at $G \geq 30.5$. So, the external periodic drive leads to a higher order periodic motion in the system. Notice also that in contrast to the case without external PD, the bifurcation diagrams do not show bistable states of the system throughout the whole range of $G$ under investigation.

\subsection{Bifurcation structure as a function of $A$}

\begin{figure}[h!]
	\centering
	\includegraphics[width=\linewidth, angle =0]{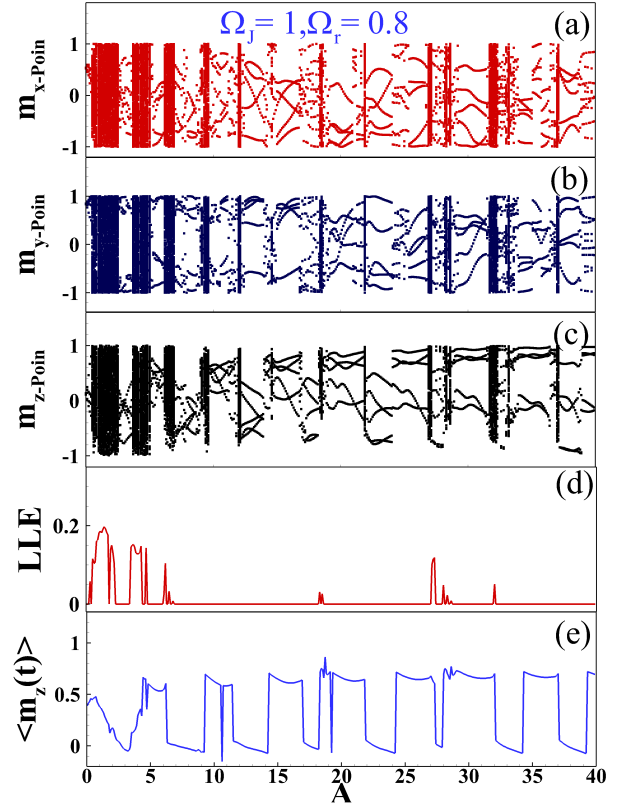}
	\caption{Bifurcation diagram of the magnetization components (a) $m_x$, (b) $m_y$, (c) $m_z$, (d) the largest Lyapunov exponent and (e) average value of $m_z-$component as functions of $A$ at $\Omega_{r} = 0.8$, $\Omega_J = 1$ and $G =5$.}
	\label{bifurc_wj_1_5_G_5_A}
\end{figure}

Significant changes in the bifurcation structure can be seen with increasing the amplitude $A$ of external PD. So, we studied the effect of $A$ as a control parameter on the bifurcation diagram which is shown in Fig.\ref{bifurc_wj_1_5_G_5_A}(a-c). The figure shows that starting from P1 motion at $A = 0, G=5$ (see Fig. \ref{bifurc_wj_1}(a-c)), the system dynamics transforms into the higher order periodic motion at $0<A\leq0.2$. Then, several chaotic bands is observed in the intervals $[0.2, 7]$ and $[27, 29]$ with a small windows of higher order periodic motion in between.

The transitions between those states manifested in the \textit{LLE} (see Fig. \ref{bifurc_wj_1_5_G_5_A}(d)) where the positive values indicate a strong chaotic response. We note also that the increase in the amplitude $A$ changes the reorientation value (see Fig. \ref{bifurc_wj_1_5_G_5_A}(e)) as it have been discussed in Ref.\cite{Kirill_2021arXiv}, but at the given simulation parameters a complete reorientation of the easy-axis does not occur.  By changing the amplitude of the external PD, one can transform the dynamics from chaotic region to higher order periodic one. Therefore, in NM-JJ system one can control the chaotic behaviour in the magnetization dynamics and reorientation process of the easy-axis.
\begin{figure}[h!]
	\centering
	\includegraphics[width=0.6\linewidth, angle =0]{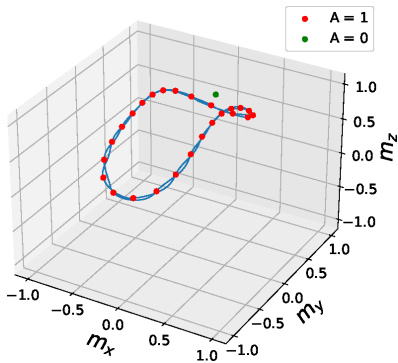}
	\caption{The Poincar\'e section (green dot) with P1 motion at $A=0$ together with the quasiperiodic orbit (blue curve) and corresponding Poincar\'e section (red dots) at $\Omega_r=0.8, A=1$. Both panels are calculated at $\Omega_J = 5$ and $G = 3$.}
	\label{poinc_wj_5}
\end{figure}

The NM-JJ system under the external PD reveals another interesting long-term behavior far from the FMR region. Namely, the quasiperiodicity is mostly appears in the weak coupling regime (at a small $G$). 
 In this case, the trajectories will never close into themselves. Fig. \ref{poinc_wj_5} demonstrates the transformation of the trajectory from the P1 motion (green dot) to the quasiperiodic one (blue curve with red dots) under the influence of PD at $G = 3$ and $\Omega_J = 5$.

\section{Conclusions}  \label{conclusion}
We have provided a detail map of various types of motions in magnetization dynamics of the nanomagnet coupled to Josephson junction. The fluctuations in the reorientation process of the easy-axis caused by the transformations between the different types of motions of the system were demonstrated. The analysis of the bifurcation diagrams revealed the exact regions where the magnetization exhibits such motions. The chaotic states, bistability, and multiperiodic orbits have been demonstrated in the resonance region. When the Josephson frequency is larger than the resonance frequency, then the bistable states and multiperiodic orbits have been observed only for the magnetization components. We found that the increase of the Josephson frequency shrinks the magnetization trajectory in space.

The chaotic behavior driven by external periodic drive have also been investigated. In this case the system shows the increase of the Josephson to magnetic energy ratio intervals of the chaotic response and the high order modes of periodic motion near the resonance. The long-term quasiperiodic behavior is manifested in the magnetization dynamics far from the resonance. In addition to this, it was found that the variation in the amplitude of the external periodic drive leads to the chaotic behavior of the system. Therefore, by applying external periodic drive one can control the dynamical behaviour of the system.

We have emphasized that the system of NM-JJ evinced nonlinear and chaotic phenomena, where a small quantitative change in the system parameters caused a huge qualitative change in the system response. Our findings can be extended to the other system of superconducting spintronics like $\varphi_{0}-$junction, which has the same current phase relation. We assume that our study will facilitate the new experimental research in this field. In particular, it might be of considerable importance for experiments on ferromagnetic resonance problems, and development of superconducting spintronics devices.

\section{Acknowledgements}

The authors are grateful to I.R. Rahmonov and A.A.Mazanik for fruitful discussion of the results of this paper. The study was carried out within the framework of the Egypt-JINR research projects. Numerical simulations were funded by the project 18-71-10095 of the Russian Scientific Fund. Special thanks to Bibliotheca Alexandrina (Egypt) and JINR (Russia) HPC for the calculating servers.

\end{document}